\documentclass[prb, twocolumn,superscriptaddress]{revtex4-1}
\usepackage{amsmath,amssymb,bm}
\usepackage{hyperref}
\usepackage{graphicx}
\usepackage{epstopdf}
\usepackage{latexsym}
\usepackage{subfigure}
\usepackage[usenames, dvipsnames]{color}
\usepackage{natbib}
\usepackage{braket}
\usepackage{float}
\usepackage[normalem]{ulem}
\usepackage{comment}
\usepackage{mathtools}
\usepackage{array}
\usepackage{tabu}

\newcommand\redsout{\bgroup\markoverwith{\textcolor{red}{\rule[0.5ex]{2pt}{0.4pt}}}\ULon}
\newcommand\bluesout{\bgroup\markoverwith{\textcolor{blue}{\rule[0.5ex]{2pt}{0.4pt}}}\ULon}

\newcommand{\ADhide}[1]{{}}
\newcommand{\SPhide}[1]{{}}

\begin{document}
\title{Comment on arxiv:1902.06475v1, 
``Magnetisation plateaus of the quantum pyrochlore Heisenberg antiferromagnet"}
\author{Sumiran Pujari}
\affiliation{Department of Physics, Indian Institute of Technology Bombay, Mumbai, MH 400076, India}
\begin{abstract}
This short note documents some of the pyrochlore magnetization plateaus
resulting from the arguments of Pal \& Lal that were not mentioned in arxiv:1902.06475v1.
These are (layered) ``Kagome"-like plateau states that are commensurate with the pyrochlore
lattice, including an exact wavefunction for the $\frac{m}{m_s}=\frac{5}{6}$
plateau state.
\end{abstract}
\maketitle

Recently, the arguments of Refs. \onlinecite{OAY}, \onlinecite{O2d}
were applied
to the Kagome \cite{Pal_Lal1} and pyrochlore \cite{Pal_Lal2} quantum Heisenberg
antiferromagnets. 
The main new ingredient was appropriate twist operators\cite{LSM}
for these highly frustrated magnets that took into account the unit cell geometry.
Using ``flux-threading" arguments, they arrived at
the following magnetization plateaus: for Kagome $\frac{m^K}{m_s}=\frac{1}{3}$ ($Q_m^K=3$),
$\frac{m^K}{m_s}=\frac{1}{9}, \frac{1}{3}, \frac{5}{9}, \frac{7}{9}$ ($Q_m^K=9$).
For pyrochlore $\frac{m^p}{m_s}=0, \frac{1}{2}$ ($Q_m^p=4$),
$\frac{m^p}{m_s}=0, \frac{1}{8}, \frac{1}{4}, \frac{3}{8}, \frac{1}{2},
\frac{5}{8}, \frac{7}{8}$ ($Q_m^p=16$). 
$m_s=\frac{1}{2}$ is
the saturation magnetization per site (in units of $\hbar$).
$Q_m$ indicates the possible enlargement
of the unit cell that often corresponds to a spontaneous translational symmetry
breaking.\cite{O2d}

However, the above pyrochlore list does not include Kagome-like states
that can be commensurately accommodated in the pyrochlore lattice.
These states may be expected to be present for the pyrochlore lattice
because of commensurability. \cite{O2d}
For Kagome lattice, the $\frac{m^K}{m_s} = \frac{7}{9}, \frac{5}{9}, 
\frac{1}{3}$ plateaus
(alternatively in terms of hardcore 
boson fillings, $\frac{1}{9}, \frac{2}{9}, \frac{3}{9}$ respectively) 
can be thought of as crystalline states in a 
$\sqrt{3} \times \sqrt{3}$ pattern (e.g. see 
Refs. \onlinecite{Nishimoto_etal2013}, \onlinecite{Okuma_etal2019}). 
This leads to the gapped, incompressible physics of the plateau states. 
For the $m=\frac{7}{9}$ plateau, exact wavefunctions can also be written 
down \cite{Schulenburg_etal2002,Zhitomirsky_Tsunetsugu2004,Bergman_etal2008,Huber_Altman2010,Changlani_PRB} 
which consist of closely-packed 
localized modes on a ($Z_3$) subset of hexagons
consistent with the $\sqrt{3} \times \sqrt{3}$ pattern. 
As has been argued in some papers, \cite{Capponi_etal2013,Nishimoto_etal2013} 
the same idea may lie behind the other plateaus, 
though exact wavefunctions have not been written down for them.

If we assume the stability of the plateaus coming from such putative 
closely-packed hexagon modes at various fillings where the sites \emph{not} 
on the hexagons are all polarized, 
then there arises a simple connection between the Kagome and pyrochlore plateaus 
because the pyrochlore lattice can be geometrically thought of 
as alternating Kagome and triangular layers connected by tetrahedral bonds. 
We can accommodate these stable closely-packed hexagon modes on the 
pyrochlore lattice in a commensurate fashion
by populating the Kagome layers with the closely-packed
hexagon modes and have the remaining sites as fully polarized \emph{including} 
those on the triangular layers. As mentioned before, commensuration gives plausibility to
their stability on the pyrochlore lattice. \cite{O2d}
Then, it remains to work out the fillings.

We first start with an exact statement.
The $\frac{m^K}{m_s}=\frac{7}{9}$ Kagome plateau ($\frac{1}{9}$ boson filling) 
corresponds
to the $\frac{m^p}{m_s}=\frac{5}{6}$ pyrochlore plateau ($\frac{1}{12}$ bosons). 
(One has three extra polarized spins from the triangular layers 
per localized hexagon on the Kagome layer.)
This is because the localized Kagome hexagon modes \emph{remain}
localized on the pyrochlore lattice too for the same (quantum interference) reasons,
\cite{Schulenburg_etal2002,Zhitomirsky_Tsunetsugu2004,Bergman_etal2008,Huber_Altman2010,Changlani_PRB}
and we can closely-pack them in the Kagome layers to get the 
$\frac{m^p}{m_s}=\frac{5}{6}$plateau.
Going further, the $\frac{m^K}{m_s}=\frac{5}{9}, \frac{1}{3}, \frac{1}{9}$ 
plateaus of Kagome will now correspond 
to $\frac{m^p}{m_s}=\frac{2}{3}, \frac{1}{2}, \frac{1}{3}$ 
pyrochlore plateaus respectively. 
Summarizing in terms of hardcore bosons, $\frac{1}{9}, \frac{2}{9}, \frac{3}{9}, \frac{4}{9}$ 
fillings on Kagome correspond
to $\frac{1}{12}, \frac{2}{12}, \frac{3}{12}, \frac{4}{12}$ 
fillings on pyrochlore respectively.

One may ask if this Kagome-pyrochlore connection
coming from the layered ``Kagome" plateaus get captured
by the arguments of Ref. \onlinecite{Pal_Lal2}, since the corresponding 
Kagome plateaus were captured by the arguments of Ref. \onlinecite{Pal_Lal1}.
Some reflection then tells us that these states indeed get captured
for $Q_m^p=12$ (not noted in Ref. \onlinecite{Pal_Lal2}). This is a tripling
of the pyrochlore unit cell, as is to be expected since in Ref. \onlinecite{Pal_Lal1},
the corresponding Kagome plateaus got captured by $Q_m^K=9$ which is 
again a tripling of the Kagome unit cell.
For Kagome, Ref. \onlinecite{Pal_Lal1} argued that 
the reason behind $Q_m^{K}$ = 3 and 9
was that $N_2$ (number of unit cells perpendicular to the flux-threading or twist
direction) be odd via Eq. 7 of Ref. \onlinecite{Pal_Lal1}. 
For pyrochlore on the other hand, the factor $4 N_2 N_3$ in 
Eq. 6 of Ref. \onlinecite{Pal_Lal2} is even owing to even number of sites 
per pyrochlore unit cell. 
Thus, $N_2$ and $N_3$ can be either odd or even without any restrictions. 
This then \textit{a priori} does not forbid $Q_m^{p}$ = 8 and 12.

For $Q_m^{p}=8$, one needs $N_2$ or $N_3$ to be even for consistency. 
This leads to the following plateaus via Eq. 6 of Ref. \onlinecite{Pal_Lal2}: 
$\frac{m^p}{m_s}=0, \frac{1}{4}, \frac{1}{2}, 
\frac{3}{4}$. This is naturally 
a subset of $Q_m^{p}$=16 plateaus mentioned in Ref. \onlinecite{Pal_Lal2}. 
The notable case is $Q_m^{p} = 12$, where one needs $N_2$ 
or $N_3$ to be a multiple of 3. This is where we get the Kagome-pyrochlore connection 
by having $N_2$ as a multiple of 3 which amounts to the tripling
of the unit cell in register with the desired Kagome-like layering. 
Then, we get the following new plateaus via Eq. 6 of Ref. \onlinecite{Pal_Lal2}: 
$\frac{m^p}{m_s}=\frac{1}{6}, \frac{1}{3}, \frac{1}{2}, 
\frac{2}{3}, \frac{5}{6}$.
The last four plateaus were arrived
at earlier in the note through heuristic arguments
by layering the Kagome plateaus for
$Q_m^{K}=9$ ($\frac{m^K}{m_s}=\frac{1}{9}, \frac{1}{3}, \frac{5}{9}, \frac{7}{9}$),
inspired by the very last case of $\frac{m^p}{m_s}=\frac{5}{6}$ 
and $\frac{m^K}{m_s}=\frac{7}{9}$ which is exact. Refs. \onlinecite{Pal_Lal1,Pal_Lal2}'s 
arguments building on 
Refs. \onlinecite{OAY,O2d} now make this heuristic reasoning into
a non-perturbative one.

Some final remarks: 
pyrochlore hosts a new plateau at $\frac{1}{6}$ 
for a tripled magnetic unit cell 
which can not be understood as a derivative of Kagome plateaus. 
What is this state?
The $\frac{1}{9}$ Kagome plateau has been numerically argued to have
no translation symmetry breaking in Ref. \onlinecite{Nishimoto_etal2013},
however taking the tripling of the unit cell 
for this plateau as argued in Ref. \onlinecite{Pal_Lal1}
at face value,
we may expect the pyrochlore-Kagome connection for this case as well.
Discussions with S. Pal and S. Lal are gratefully acknowledged.

\bibliographystyle{apsrev4-1}
\bibliography{biblio}
\end{document}